\documentstyle[12pt,epsf]{article}
\renewcommand{\baselinestretch}{1.2}
\newcommand{\uno}{{ 1\:\!\!\!\mbox{I}}}

\def\npb#1#2#3{    {\it Nucl. Phys. }{\bf B #1} (19#2) #3}
\def\plb#1#2#3{    {\it Phys. Lett. }{\bf B #1} (19#2) #3}
\def\prd#1#2#3{    {\it Phys. Rev. }{\bf D #1} (19#2) #3}

\def\prl#1#2#3{    {\it Phys. Rev. Lett. }{\bf #1} (19#2) #3}

\def\ppnp#1#2#3{   {\it Prog. Part. Nucl. Phys. }{\bf #1} (19#2) #3}

\def\zpc#1#2#3{    {\it Zeit. f\"ur Physik }{\bf C #1} (19#2) #3}

\def\eq#1{{eq.~(\ref{#1})}}

\newcommand{\bea}{\begin{eqnarray}}
\newcommand{\beq}{\begin{equation}}
\newcommand{\eea}{\end{eqnarray}}
\newcommand{\eeq}{\end{equation}}

\newcommand{\diag}{{\scriptscriptstyle{\rm{diag}}}}

\begin{document}
\begin{flushright}
Rome1-1113/95\\
ROM2F/95/23
\end{flushright}
\vskip 2.0cm
\centerline{\Large{\bf{Constraints on supersymmetry from}}}
\vspace*{1.0ex}
\centerline{\Large{\bf{FCNC and CP violation}}}
\vspace*{6.0ex}
\centerline{\large{ E. Gabrielli}}
\vspace*{1.5ex}
\centerline{\it Dip. di Fisica,
Universit\`a di Roma ``La Sapienza" and}
\centerline{\it INFN, Sezione di Roma,
P.le A. Moro 2, I-00185 Roma, Italy. }
\vspace*{1.5ex}
\centerline{\large{ A. Masiero$^{*}$}}
\vspace*{1.5ex}
\centerline{\it Dip. di Fisica, Universit\`{a} di Perugia and}
\centerline{\it INFN, Sezione di Perugia,
Via Pascoli, I-06100 Perugia, Italy.}
\vspace*{1.5ex}
\centerline{\large{ L. Silvestrini}}
\vspace*{1.5ex}
\centerline{\it  Dip. di Fisica, Univ. di Roma ``Tor Vergata''
and INFN, Sezione di Roma II,}
\centerline{\it Via della Ricerca Scientifica 1, I-00133 Roma, Italy.}
\begin{abstract}
We consider FCNC and CP violating processes mediated by gluino exchange in
generalized supersymmetric theories.
We present the constraints on flavour (F) changing squark mass terms at the
electroweak energy scale, focusing our analysis on $\Delta F = 2$ transitions.
Results are also given for $\Delta F = 1$ CP conserving processes both in the
hadronic and leptonic sectors, while we limit ourselves to some remarks on the
relevance of box diagrams in the evaluation of $\epsilon^{\prime}/\epsilon$.
\end{abstract}
\vspace*{3.0ex}
\centerline{$^{*}$ Talk given by A. Masiero at SUSY'95, Ecole Polytechnique,
Palaiseau, May 1995.}

\vskip 3.0cm

\vfill\eject
\pagestyle{empty}\clearpage
\newpage
Flavour changing neutral currents (FCNC) and CP violating processes provide us
with a powerful tool to constrain the mass spectrum of supersymmetric
particles \cite{susy2}-\cite{susycp}.
Lately, there has been a renewed interest in the analysis of these
processes, in connection with the study of supersymmetric grand-unified
theories (SUSY-GUTs) \cite{susyguts} and of supersymmetry
(SUSY) breaking in effective
supergravities which emerge as the low energy limit of superstring theories
\cite{string}.
Indeed, from the FCNC tests one can obtain relevant constraints
on the mechanism of SUSY breaking and on the
interactions of fermions up to the supergravity breaking scale.
One very efficient way to obtain such constraints is to analyse
gluino-mediated FCNC processes, which can be sizeable due to the presence
of the strong coupling \cite{susyfcnc,susycp}.
These flavour changing (FC) transitions arise if the
squark mass matrices are not diagonalizable with the same transformations
which diagonalize quark mass matrices.
In this case a convenient choice is to
keep the gluino-quark-squark ($\tilde{g}-q-\tilde{q}$) vertices diagonal in
flavour space. This can be obtained by applying to the squark fields
exactly the same
transformations that diagonalize the quark mass matrices.
In this basis, usually called the super-KM one, all flavour changing effects
are due to off-diagonal mass insertions in squark propagators \cite{hall}.
As long as the ratio of the off-diagonal entries over an average squark mass
remains a small parameter, the first term in the expansion obtained by an
off-diagonal mass insertion represents a suitable approximation.
The advantage of this method is that it avoids the specific knowledge of the
sfermion mass matrices, and simplifies the phenomenological analysis.

The presence of off-diagonal mass terms at the electroweak scale
can be due to the initial conditions: SUSY breaking terms may yield
contributions which are not universal, i.e. they are not proportional to the
unit matrix in flavour space.
Otherwise, even starting with universal mass contribution
to sfermions in the SUSY soft breaking sector, renormalization effects from
the starting point, i.e. the scale of supergravity breaking,
down to the Fermi scale
can bring about a misalignment between $q$ and $\tilde{q}$ mass matrices
\cite{susyfcnc,susycp}.
This latter situation is what we encounter in the minimal SUSY standard model.
For instance, consider the mass matrix squared
of the scalar partner of the left-handed down-quarks $d_L$.
At the scale of supergravity
breaking this matrix consists of the SUSY conserving contribution
$m_dm_d^{\dagger}$ (where $m_d$ denotes the down
quark mass matrix) and the SUSY
breaking universal contribution $\tilde{m}^2{\uno}$.
However, the term $h_uQHu^c$ of the superpotential
generates a logarithmically
divergent contribution which is proportional to
$h_uh_u^{\dagger}$ and, hence,
to $m_um_u^{\dagger}$ ($m_u$ being the up-quark mass matrix).
Hence the resulting $\tilde{d}_L$ mass matrix squared at the Fermi scale is:
\beq
m^2_{\tilde{d}_L\tilde{d}_L}=m_dm^{\dagger}_d+
\tilde{m}^2 {\uno} +c\,m_um^{\dagger}_u
\label{msdown}
\eeq
Switching now to the super-KM basis, rotating the $\tilde{q}$ fields together
with the $q$ ones, we obtain off-diagonal mass terms due to the presence of
the last term in eq. (\ref{msdown}).
{}From \eq{msdown} it is easy to realize that the
mass insertion needed to accomplish the transition from $\tilde{d}_{iL}$
to $\tilde{d}_{jL}$ ($i,j$ flavour indices) is given by:
\beq
\left(\Delta_{LL}^d\right)_{ij}=
c\left[K\left(m_u^{\diag}
\right)^2K^{\dagger}\right]_{ij}
\label{massins}
\eeq
where $K$ is the Cabibbo-Kobayashi-Maskawa matrix and
$m_u^{\diag}$
denotes the diagonalized up-quark mass matrix.
In the following, with the notation
$\left(\Delta_{AB}^q\right)_{ij}$, we mean the mass insertion needed
for a transition from a squark $\tilde{q}_{iA}$ to $\tilde{q}_{jB}$
with $A=(L,R)$ and $B=(L,R)$.
Actually, as particularly
emphasized in ref. \cite{hagelin}, the above expression for
$(\Delta_{LL}^{d})_{ij}$
may be somewhat misleading since one might think that the term $c$
in the r.h.s. of \eq{massins} is a constant
(i.e. independent of the SUSY breaking scale) or at most depends
logarithmically on it. On the contrary, the one-loop RGE's show that $c$
depends
quadratically on that scale. Since also the average squark mass is proportional
to this scale, the meaningful parameter for our mass insertion approximation is
the dimensionless quantity $\delta=\Delta/\tilde{m}^2$,
where $\tilde{m}$ denotes at the same time the average squark mass
and the typical SUSY breaking scale. This observation is of utmost
relevance if one wants to understand the scaling of the SUSY contribution
to FCNC with increasing squark masses. The powers of $\tilde{m}$ in the
denominator which are present to compensate for $\Delta$ mass insertions
in the numerator do not have to be considered if also $\Delta$ is proportional
to $\tilde{m}^2$. This justifies why gluino-induced FCNC SUSY contributions
remain still sizeable even for $\tilde{q}$ masses above 1 TeV \cite{hagelin},
as we will see in what follows.
Two peculiar features of the MSSM should be noted.
The first is that in this model there is a sharp hierarchy among the
$\Delta_{AB}$: $(\Delta_{LL})_{ij}>>(\Delta_{LR})_{ij}>>(\Delta_{RR})_{ij}$
with
$i \neq j$, due to the different number of mass insertions needed to
accomplish the various transitions.
The second peculiar feature is the smallness of these terms due to the super-
GIM mechanism. We stress that there is no reason in the general case to expect
such a pattern of FC effects.

There exist three main analysis of the constraints on $\delta$'s in the
literature: ref. \cite{gabbmas,hagelin,gms}. In particular, in \cite{gms} some
previous discrepancies are discussed and more emphasis is provided on the
constraints on the imaginary parts of the $\delta$'s given by CP violation.

We now come to the results of our analysis concerning the terms
$(\Delta_{LL})_{ij}$, $(\Delta_{LR})_{ij}$ and $(\Delta_{RR})_{ij}$
in the u- and d-sectors. In the following we consider the case in which
$(\Delta_{LR})_{ij}\simeq (\Delta_{RL})_{ij}$.
We will comment later on the analogous contributions
in the charged lepton sector.

\begin{figure}
\epsfysize=8.0cm
\epsfxsize=17.0cm
\epsffile{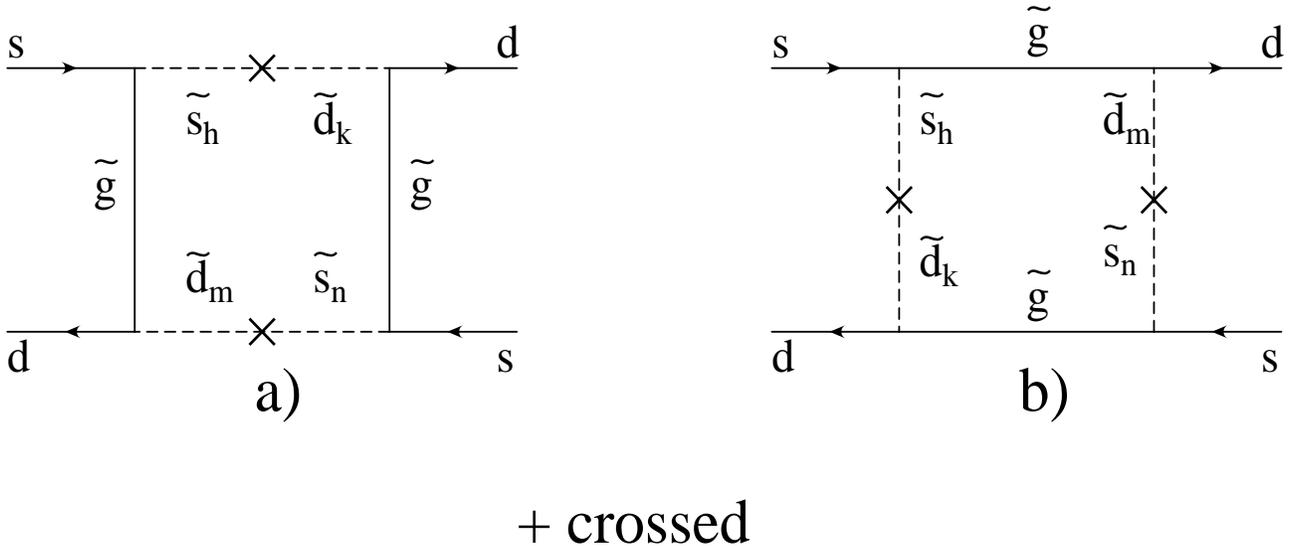}
\caption{The gluino box diagrams for $\Delta S=2$ transitions.}
\label{boxes}
\end{figure}

First we consider $\Delta F =2$ FCNC processes.
In the down sector the
$\Delta_{ij}$ mass insertions are bounded by the $K-\bar{K}$
mass difference and by the CP violating parameter $\epsilon$
$(\delta_{12})$ and the $B_d-\bar{B}_d$ mixing $(\delta_{13})$,
while the only available bound in the up-sector concerns
$\delta_{12}$ from $D-\bar{D}$ mixing.

\begin{figure}   
    \begin{center}
    \epsfysize=9.5truecm
    \leavevmode\epsffile{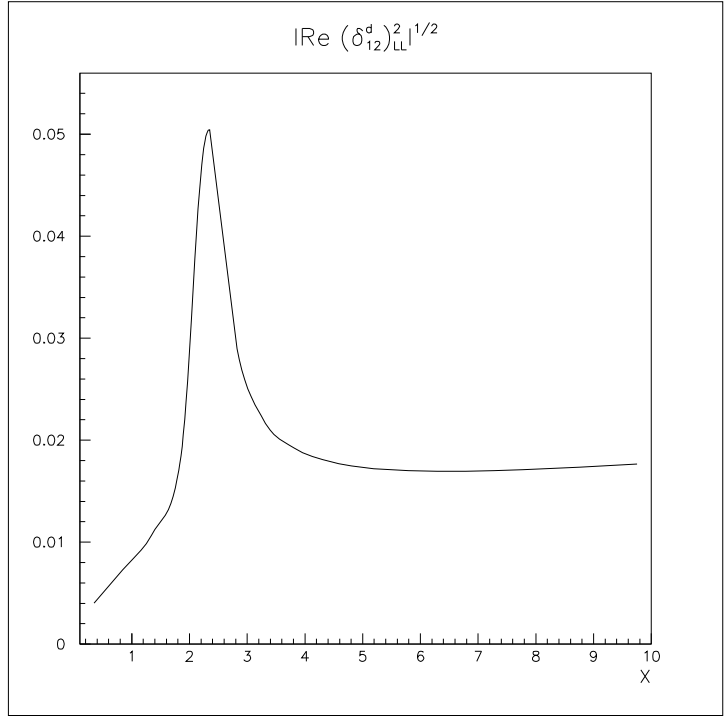}
    \end{center}
    \caption[]{The $\sqrt{\left|{\mbox Re}  \left(\delta^{d}_{12}
\right)_{LL}^{2}\right|} $ as a function
     of $x=m_{\tilde{g}}^2/\tilde{m}^2$, for  a squark mass
     $\tilde{m}=100\mbox{GeV}$.}
\label{kkll}
\end{figure}

\begin{figure}[t]   
    \begin{center}
    \epsfysize=9.5truecm
    \leavevmode\epsffile{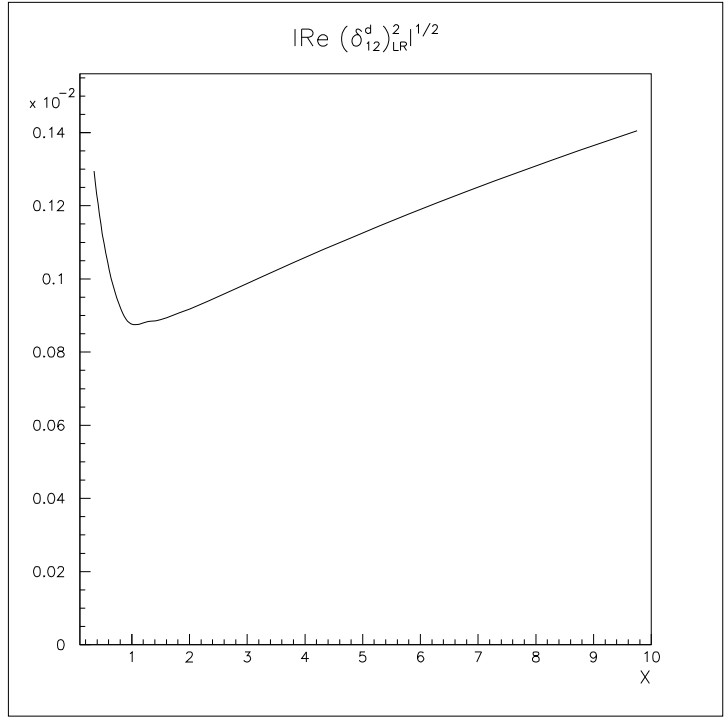}
    \end{center}
    \caption[]{The $\sqrt{\left|{\mbox Re}  \left(\delta^{d}_{12}
\right)_{LR}^{2}\right|} $ as a function
     of $x=m_{\tilde{g}}^2/\tilde{m}^2$, for  a squark mass
     $\tilde{m}=100\mbox{GeV}$.}
\label{kklr}
\end{figure}

\begin{figure}[b]   
    \begin{center}
    \epsfysize=9.5truecm
    \leavevmode\epsffile{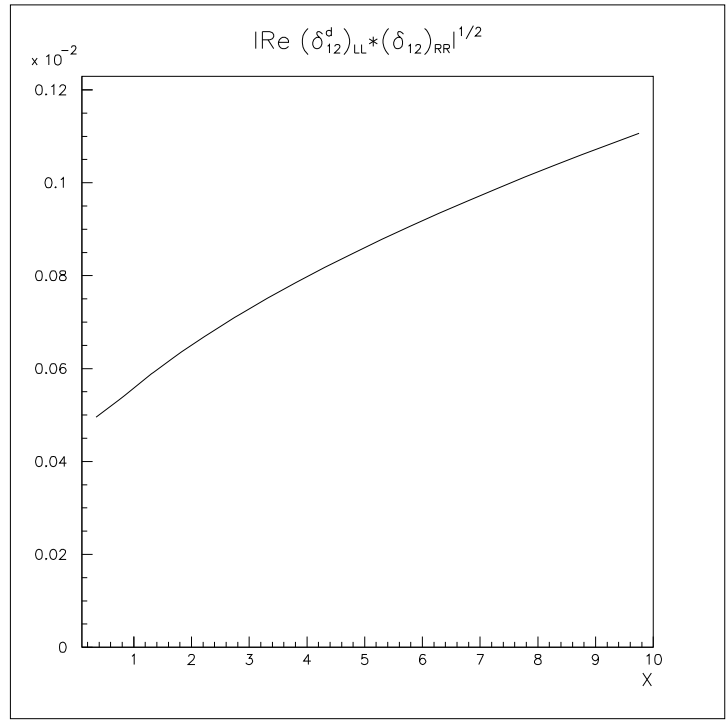}
    \end{center}
    \caption[]{The $\sqrt{\left|{\mbox Re}  \left(\delta^{d}_{12}
\right)_{LL} \left(\delta^{d}_{12}
\right)_{RR}\right|} $ as a function
     of $x=m_{\tilde{g}}^2/\tilde{m}^2$, for  a squark mass
     $\tilde{m}=100\mbox{GeV}$.}
\label{kkllrr}
\end{figure}

The effective hamiltonian for $\Delta S = 2$ processes can be obtained from
the calculation of the diagrams in fig. \ref{boxes}.
We give below the mass difference $\Delta m_K$, obtained by considering the
matrix element of the effective hamiltonian between Kaon states:
\begin{eqnarray}
\Delta m_K &=& \frac{\alpha_s^2 f_k^2 m_K}{81 \tilde{m}^2} \, \,
\left\{ \left[\left(\delta^d_{12}\right)^2_{LL} + \left(\delta^d_{12}\right)
^2_{RR}\right] \, \left(-6xM(x)+11G(x)\right)
\right. \nonumber  \\
&+& \left[\left(\delta^d_{12}\right)^2_{LR} + \left(\delta^d_{12}\right)
^2_{RL}\right]\, 33Z\,xM(x) \nonumber \\
&+& \left(\delta^d_{12}\right)^2_{LR}\left(\delta^d_{12}\right)^2_{RL} \,
\left( -24 Z - 14 \right)G(x) \nonumber \\
&+& \left. \left(\delta^d_{12}\right)^2_{LL}\left(\delta^d_{12}\right)^2_{RR}
\,
\left[\left( -96 Z - 18\right)\,xM(x)  -\left(4Z - 6\right)\,G(x) \right]
\right\}
\label{dmk}
\end{eqnarray}
where we have denoted by $Z$ the quantity
\begin{equation}
Z=\frac{m_K^2}{\left( m_s + m_d\right)^2}
\end{equation}
and the functions $M(x)$ and $G(x)$ can be found in ref. \cite{gabbmas}.
Concerning this calculation we find some discrepancies with ref.
\cite{gabbmas,hagelin}. More details on the subject will be provided in
ref. \cite{ggms}.
Our numerical results on the limits obtained from $\Delta F=2$ processes are
shown in Tables \ref{reds2} and \ref{imdelta12}, and in figures \ref{kkll},
 \ref{kklr}, \ref{kkllrr}.

 \begin{table}
 \begin{center}
 \begin{tabular}{||c|c|c|c||}  \hline \hline
 $x$ & $\sqrt{\left|{\mbox Re}  \left(\delta^{d}_{12} \right)_{LL}^{2}\right|}
$
 &
 $\sqrt{\left|{\mbox Re}  \left(\delta^{d}_{12} \right)_{LR}^{2}\right|} $ &
 $\sqrt{\left|{\mbox Re}  \left(\delta^{d}_{12}
\right)_{LL}\left(\delta^{d}_{12}
 \right)_{RR}\right|} $ \\
 \hline
 $
   0.3
 $ &
 $
1.9\times 10^{-2}
 $ & $
7.9\times 10^{-3}
 $ & $
2.5\times 10^{-3}
 $ \\
 $
   1.0
 $ &
 $
4.0\times 10^{-2}
 $ & $
4.4\times 10^{-3}
 $ & $
2.8\times 10^{-3}
 $ \\
 $
   4.0
 $ &
 $
9.3\times 10^{-2}
 $ & $
5.3\times 10^{-3}
 $ & $
4.0\times 10^{-3}
 $ \\ \hline \hline
 $x$ & $\sqrt{\left|{\mbox Re}  \left(\delta^{d}_{13} \right)_{LL}^{2}\right|}
$
 &
 $\sqrt{\left|{\mbox Re}  \left(\delta^{d}_{13} \right)_{LR}^{2}\right|} $ &
 $\sqrt{\left|{\mbox Re}  \left(\delta^{d}_{13}
\right)_{LL}\left(\delta^{d}_{13}
 \right)_{RR}\right|} $ \\
 \hline
 $
   0.3
 $ &
 $
4.6\times 10^{-2}
 $ & $
5.6\times 10^{-2}
 $ & $
1.6\times 10^{-2}
 $ \\
 $
   1.0
 $ &
 $
9.8\times 10^{-2}
 $ & $
3.3\times 10^{-2}
 $ & $
1.8\times 10^{-2}
 $ \\
 $
   4.0
 $ &
 $
2.3\times 10^{-1}
 $ & $
3.6\times 10^{-2}
 $ & $
2.5\times 10^{-2}
 $ \\ \hline \hline
 $x$ & $\sqrt{\left|{\mbox Re}  \left(\delta^{u}_{12} \right)_{LL}^{2}\right|}
$
 &
 $\sqrt{\left|{\mbox Re}  \left(\delta^{u}_{12} \right)_{LR}^{2}\right|} $ &
 $\sqrt{\left|{\mbox Re}  \left(\delta^{u}_{12}
\right)_{LL}\left(\delta^{u}_{12}
 \right)_{RR}\right|} $ \\
 \hline
 $
   0.3
 $ &
 $
4.7\times 10^{-2}
 $ & $
6.3\times 10^{-2}
 $ & $
1.6\times 10^{-2}
 $ \\
 $
   1.0
 $ &
 $
1.0\times 10^{-1}
 $ & $
3.1\times 10^{-2}
 $ & $
1.7\times 10^{-2}
 $ \\
 $
   4.0
 $ &
 $
2.4\times 10^{-1}
 $ & $
3.5\times 10^{-2}
 $ & $
2.5\times 10^{-2}
 $ \\ \hline \hline
 \end{tabular}
 \caption[]{Limits on $\mbox{Re}\left(\delta_{ij}\right)_{AB}\left(
\delta_{ij}\right)_{CD}$, with $A,B,C,D=(L,R)$, for a squark mass
 $\tilde{m}=500\mbox{GeV}$ and for different values of
 $x=m_{\tilde{g}}^2/\tilde{m}^2$.}
 \label{reds2}
 \end{center}
 \end{table}
 \begin{table}
 \begin{center}
 \begin{tabular}{||c|c|c|c||}  \hline \hline
  $x$ &
 ${\scriptstyle\sqrt{\left|{\mbox Im}  \left(\delta^{d}_{12} \right)_{LL}^{2}
\right|} }$ &
 ${\scriptstyle\sqrt{\left|{\mbox Im}  \left(\delta^{d}_{12} \right)_{LR}^{2}
\right|} }$ &
 ${\scriptstyle\sqrt{\left|{\mbox Im}  \left(\delta^{d}_{12}
\right)_{LL}\left(\delta^{d}_{12}
 \right)_{RR}\right|} }$ \\
 \hline
 $
   0.3
 $ &
 $
1.5\times 10^{-3}
 $ & $
6.3\times 10^{-4}
 $ & $
2.0\times 10^{-4}
 $ \\
 $
   1.0
 $ &
 $
3.2\times 10^{-3}
 $ & $
3.5\times 10^{-4}
 $ & $
2.2\times 10^{-4}
 $ \\
 $
   4.0
 $ &
 $
7.5\times 10^{-3}
 $ & $
4.2\times 10^{-4}
 $ & $
3.2\times 10^{-4}
 $ \\ \hline \hline
 \end{tabular}
 \ \caption[]{Limits on
$\mbox{Im}\left(\delta_{12}^{d}\right)_{AB}\left(\delta_{12}^{d}\right)_{CD}$,
with $A,B,C,D=(L,R)$, for
 a squark mass $\tilde{m}=500\mbox{GeV}$ and for different values of
 $x=m_{\tilde{g}}^2/\tilde{m}^2$.}
 \label{imdelta12}
 \end{center}
 \end{table}

Let us now turn to $\Delta F=1$ FCNC processes. We obtain limits from $b \to s
\gamma $ decays ($\delta^{d}_{23}$) in the squark sector.
Numerical results are given in table \ref{bsg}.
 \begin{table}
 \begin{center}
 \begin{tabular}{||c|c|c||}  \hline \hline
  & & \\
 $x$ & $\left|\left(\delta^{d}_{23} \right)_{LL}\right| $ &
 $\left|  \left(\delta^{d}_{23} \right)_{LR}\right| $ \\
  & & \\ \hline
 $
   0.3
 $ &
 $
4.4
 $ & $
1.3\times 10^{-2}
 $ \\
 $
   1.0
 $ &
 $
8.2
 $ & $
1.6\times 10^{-2}
 $ \\
 $
   4.0
 $ &
 $
26
 $ & $
3.0\times 10^{-2}
 $ \\ \hline \hline
 \end{tabular}
 \caption[]{Limits on the $\left| \delta_{23}^{d}\right|$ from
 $b\rightarrow s \gamma$ decay for a squark mass $\tilde{m}=500\mbox{GeV}$
 and for different values of $x=m_{\tilde{g}}^2/\tilde{m}^2$.}
 \label{bsg}
 \end{center}
 \end{table}

Table \ref{bsg} shows that the decay
$(b\rightarrow s+\gamma)$ does not limit the $\delta_{LL}$ insertion
for a SUSY breaking of O(500 GeV). Indeed, even taking
$m_{\tilde{q}}=100\mbox{GeV}$, the term $(\delta_{23})_{LL}$ is only marginally
limited ( $(\Delta_{LL})_{23}<0.3$ for $x=1$).
Obviously, $(\delta_{23}^d)_{LR}$
is much more constrained since with a $\Delta_{LR}$ FC mass insertion the
helicity flip needed for $(b\rightarrow s+\gamma)$ is realized in the gluino
internal line and so this contribution has an amplitude enhancement
of a factor $m_{\tilde{g}}/m_b$ over the previous case with
$\Delta_{LL}$.

A similar analysis can be performed in the leptonic sector where the masses
$\tilde{m}$ and $m_{\tilde{g}}$ are replaced by the average slepton
mass and the photino mass $m_{\tilde{\gamma}}$ respectively.
A clear but important point to be stressed is that the
severe bounds that we provide
on the $\delta_{LL}$ and $\delta_{LR}$  mass insertions in the leptonic
sector and the consequent need for high degeneracy of charged sleptons,
only apply if separate lepton numbers are violated. It is well known
that in the MSSM the lepton numbers $L_{e},~L_{\mu}$ and $L_{\tau}$
are separately conserved because of the diagonality of the soft
breaking terms and the masslessness of neutrinos.
If at least one of this two properties is not present one can have
partial lepton number violation. A particularly interesting example is
the case where neutrinos acquire a mass through a see-saw mechanism
(for its SUSY version and the implications for FCNC see \cite{borzmas}).
\begin{table}
 \begin{center}
 \begin{tabular}{||c|c|c||}  \hline \hline
  & & \\
 $x$ & $\left|\left(\delta^{l}_{12} \right)_{LL}\right| $ &
 $\left|  \left(\delta^{l}_{12} \right)_{LR}\right| $ \\
  & & \\ \hline
 $
   0.3
 $ &
 $
4.1\times 10^{-3}
 $ & $
1.4\times 10^{-6}
 $ \\
 $
   1.0
 $ &
 $
7.7\times 10^{-3}
 $ & $
1.7\times 10^{-6}
 $ \\
 $
   5.0
 $ &
 $
3.2\times 10^{-2}
 $ & $
3.8\times 10^{-6}
 $ \\ \hline \hline
  & & \\
 $x$ & $\left|\left(\delta^{l}_{13} \right)_{LL}\right| $ &
 $\left|  \left(\delta^{l}_{13} \right)_{LR}\right| $ \\
  & & \\ \hline
 $
   0.3
 $ &
 $
15
 $ & $
8.9\times 10^{-2}
 $ \\
 $
   1.0
 $ &
 $
29
 $ & $
1.1\times 10^{-1}
 $ \\
 $
   5.0
 $ &
 $
1.2\times 10^{2}
 $ & $
2.4\times 10^{-1}
 $ \\ \hline \hline
  & & \\
 $x$ & $\left|\left(\delta^{l}_{23} \right)_{LL}\right| $ &
 $\left|  \left(\delta^{l}_{23} \right)_{LR}\right| $ \\
  & & \\ \hline
 $
   0.3
 $ &
 $
2.8
 $ & $
1.7\times 10^{-2}
 $ \\
 $
   1.0
 $ &
 $
5.3
 $ & $
2.0\times 10^{-2}
 $ \\
 $
   5.0
 $ &
 $
22
 $ & $
4.4\times 10^{-2}
 $ \\ \hline \hline
 \end{tabular}
 \caption[]{Limits on the $\left| \delta_{ij}^{d}\right|$ from
 $l_j\rightarrow l_i \gamma$ lepton decay for
 a slepton mass $\tilde{m}=100\mbox{GeV}$ and for different values of
 $x=m_{\tilde{\gamma}}^2/\tilde{m}^2$.}
 \label{deltal}
 \end{center}
 \end{table}
In table \ref{deltal} we exhibit the bounds on $\delta_{LL}^l$ and
$\delta_{LR}^l$
coming from the limits on
$\mu\rightarrow e\gamma,~\tau\rightarrow e\gamma$ and
$\tau\rightarrow \mu\gamma$, for a slepton mass of O(100 GeV)
and for different values of $x=m_{\tilde{\gamma}}^2/\tilde{m}^2$.
Our results confirm those obtained in refs \cite{gabbmas,hagelin}.

  Finally we make a short comment on the gluino-induced FC contribution
to $\epsilon^{\prime}/\epsilon$ (the interested reader should consult ref.
\cite{gms} for a more thorough discussion).
A common feature of the past literature
on this point was the emphasis on the role of superpenguins, while
superbox contributions were thought to be negligible. In our recent work
\cite{gms} we show that box contributions yield a sizeable interference effect
with the superpenguins ones. The results concerning the bounds on the
imaginary parts of $(\delta^d_{12})_{LL}$ and $(\delta^d_{12})_{LR}$
from $\epsilon^{\prime}/\epsilon$ for an
average squark mass of 500 GeV are provided in table \ref{epp}.
It is apparent from a comparison of table \ref{imdelta12} and table \ref{epp}
that SUSY models
with predominantly $\delta_{LL}$ or $\delta_{RR}$ contributions to CP
violation
(like the MSSM) tend to be superweak, while models with sizeable
$\delta_{LR}$ are likely to be milliweak. Complete expressions for the separate
box and penguin contributions to the $\Delta S=1$ effective hamiltonian
will be provided in a forthcoming paper \cite{ggms}.
   The implications of the constraints obtained in this work on SUSY-GUTs
and on theories with non-universal soft breaking terms are presently
under study \cite{ggms}.

 \begin{table}
 \begin{center}
 \begin{tabular}{||c|c|c||}  \hline \hline
  $x$ & ${\scriptstyle\left|{\mbox Im} \left(\delta^{d}_{12}  \right)_{LL}
\right|} $ &
 ${\scriptstyle\left|{\mbox Im} \left(\delta^{d}_{12}\right)  _{LR}\right| }$
\\\hline
 $
   0.3
 $ &
 $
1.0\times 10^{-1}
 $ & $
1.1\times 10^{-5}
 $ \\
 $
   1.0
 $ &
 $
4.8\times 10^{-1}
 $ & $
2.0\times 10^{-5}
 $ \\
 $
   4.0
 $ &
 $
2.6\times 10^{-1}
 $ & $
6.3\times 10^{-5}
 $ \\ \hline \hline
 \end{tabular}
 \caption[]{Limits from $\epsilon^{\prime}/\epsilon$ on
$\mbox{Im}\left(\delta_{12}^{d}\right)$, for
 a squark mass $\tilde{m}=500\mbox{GeV}$ and for different values of
 $x=m_{\tilde{g}}^2/\tilde{m}^2$.}
 \label{epp}
 \end{center}
 \end{table}

\renewcommand{\baselinestretch}{1}

\newpage

\end{document}